\documentclass[preprint,showpacs,superscriptaddress, amsmath,amssymb]{revtex4-2}
\usepackage[utf8]{inputenc}
\usepackage{color}
\usepackage{comment}
\usepackage{natbib}
\usepackage{graphicx}
\usepackage{hyperref}
\usepackage{cleveref}
\usepackage{bm}
\usepackage{amssymb}
\usepackage{siunitx}
\usepackage[normalem]{ulem}

\definecolor{darkgreen}{rgb}{0.0, 0.5, 0.0}
\newcommand{\ea}[1]{\textcolor{darkgreen}{#1}}    % for Emilien
\begin{document}

%\title{Experimental compaction of soft spherical packings: A micromechanical study}
%\title{ Illuminating the scales in highly soft granular materials compressed far after the jammed state using advanced 3D digital image correlation (DIC)}
%\title{Exploring the scales in highly soft grain assemblies compressed far beyond the jammed state using advanced 3D digital image correlation}
%\title{Exploring the scales in highly soft grain assemblies using advanced 3D digital image correlation}

%\title{Multiscale exploration when compacting soft spherical packings far beyond the jammed state : A 3D micromechanical study}
\title{Compacting an assembly of soft balls far beyond the jammed state: insights from 3D imaging}
%%EA --> hummm il ne manque pas le mot "experimental" ou "l'idee que c'est experimental" quelque part ?

\author{Jonathan Bar\'es}\email{jonathan.bares@umontpellier.fr}
\affiliation{LMGC, Universit\'e de Montpellier, CNRS, Montpellier, France}
\author{Manuel Cárdenas-Barrantes}\email{macardenasb@fc.ul.pt}
\affiliation{CFTC, Universidade de Lisboa, Lisbonne, Portugal}
\author{Gustavo Pinz\'on}\email{gustavo.pinzon@3sr-grenoble.fr}
\affiliation{Universit\'e Grenoble Alpes, Grenoble INP, CNRS, 3SR, F-38000 Grenoble, France}
\author{Edward And\`o}\email{edward.ando@epfl.ch}
\affiliation{EPFL Center for Imaging, École Polytechnique Fédérale de Lausanne (EPFL)}
\author{Mathieu Renouf}\email{mathieu.renouf@umontpellier.fr}
\affiliation{LMGC, Universit\'e de Montpellier, CNRS, Montpellier, France}
\author{Gioacchino Viggiani}\email{cino.viggiani@3sr-grenoble.fr}
\affiliation{Universit\'e Grenoble Alpes, Grenoble INP, CNRS, 3SR, F-38000 Grenoble, France}
\author{Emilien Az\'ema}\email{emilien.azema@umontpellier.fr}
\affiliation{LMGC, Universit\'e de Montpellier, CNRS, Montpellier, France}
\affiliation{Institut Universitaire de France (IUF), Paris, France}

\date{\today}

\begin{abstract}
Very soft grain assemblies have unique shape-changing capabilities that allow them to be compressed far beyond the rigid jammed state by filling void spaces more effectively. However, accurately following the formation of these systems by monitoring the creation of new contacts, the changes in grain shape, and measuring grain-scale stresses is challenging. We developed an experimental method overtaking these challenges and connecting their microscale behavior to their macroscopic response. By tracking the local strain energy during compression, we reveal a transition from granular-like to continuous-like material. Mean contact geometry is shown to vary linearly with the packing fraction, which is supported by a mean field approximation. We also validate a theoretical framework which describes the compaction from a local view. Our experimental framework provides insights into the granular micro-mechanisms and opens new perspectives for rheological analysis of highly deformable grain assemblies in various fields ranging from biology to engineering. %[149 words]
%EA --> YES !!
%MC --> OUI

\end{abstract}

\maketitle

%/<|>\%/<|>\%/<|>\%/<|>\%/<|>\%/<|>\%/<|>\%/<|>\%/<|>\%/<|>\%/<|>\%/<|>\%
% intro
\section{Introduction}

%Situation
Compressing a disordered assembly of grains is a seemingly simple process that has been carried out since the dawn of time by humanity to store, transport or transform matter. The objectives are quite basic: either to improve the strength of the resulting material, or to optimize the space by reducing the volume taken by the grains.
In the most common case of hard grains (\textit{i.e.}, grains with high stiffness relatively to the applied pressure, $P$), the compression generally ends once the grains cannot rearrange any more, meaning a mechanically stable state is reached unless grains break.
This so called jammed state \cite{liu1998_nat,bi2011_nat,behringer2018_rpp}, mainly depends on the grain morphologies (size \cite{nguyen2014effect} and shape \cite{Hecke2010,kyrylyuk2011effect}), the friction between them \cite{Hecke2010}, and the loading process \cite{bi2011_nat}. 
The packing fraction $\phi$ cannot exceed that of the Random Close Packing (RCP), and the coordination number $Z$ (\textit{i.e.}, the average number of contacts per grain) cannot be higher than $6$ for spherical ($12$ for non-spherical) grain assemblies. 
It is now well documented that the applied stresses are transmitted by an inhomogeneous network of forces where only a small proportion of grains bear the strongest forces \cite{radjai1996_prl,peters2005_pre}. 
The elastic properties of the assembly, around the jammed state, cannot be directly deduced from grains elasticity, but result from a subtle intertwining between the values of the packing fraction, the coordination number and contact properties \cite{Makse1999,Agnolin2007c,Khalili2017b}. The same is true for the coordination number, which evolves as a power law of the packing fraction \cite{durian1995_prl,majmudar2007_prl,katgert2010_epl}.

But, what if we consider soft and even squishy grains \cite{bares2022_pp} (\textit{i.e.}, grains having a low stiffness relatively to the applied pressure) while continuing to compress the packing?
The grains would deform; changing their shapes elastically or in a more complex manner, without breaking to accommodate for the internal mechanical and geometrical constraints. As a result, the solid fraction would exceed that of the RCP, and it would even be possible to entirely fill the space. 
A new world, yet poorly understood, is opening up with these materials made of soft, squishy, deformable grains. 
Such materials play a major role in various fields of natural sciences (foams, colloidal suspensions, biological cells, blood clogging in a vein, \textit{etc}) and in many industrial processes (emulsions, metallic powders, sintered material, rubber mixtures, \textit{etc}). In soft granular assemblies, the macro- and micro- scale behaviors are no longer uniquely ruled by the steric exclusions and the topological disorder of contacts, but also by the bulk behavior of each constitutive grain.
Including the change in grain shape in addition to the inherently multi-contact nature of a granular system in a realistic 3D modeling is a vast emerging topic, with challenging experimental, numerical, and theoretical issues \cite{bares2022_pp,cantor2022_pp}.

From a numerical point of view, these last years have seen the advent of new strategies in the form of Discrete Element Methods (DEM) coupled with different continuum approaches \cite{bares2022_pp}. 
Although technical challenges still remain, especially in terms of computational time, existing studies, mainly in 2D and few in 3D \cite{HARTHONG2012784,Ku_compact_def_2023,ABDELMOULA2017142}, have allowed to glimpse new microscopic mechanisms that take place from contacts scale until the inside of the grains. For example, the power law dependence of the coordination number with the packing fraction is still valid far beyond the jammed state \cite{vu2019_pre,Nezamabadi2019,cantor2020_prl,cardenas2022_sm,cardenas2022_pre}, even for assemblies of initially non-circular grains \cite{cardenas2021_pre}. 
Local grain rearrangements, sometimes dramatic, can still occur even after the jamming point \cite{howell1999_prl,daniels2008_jgr,lebouil2014_prl,bares2017_pre}, but at the same time the stress distribution becomes more homogeneous as the packing fraction increases. Friction also contributes to a better homogenization of the contact force network \cite{cardenas2021_pre}. 
By implementing some of these new microscopic descriptors into a rigorous micro-mechanical framework, Cantor {\it et al.} (in 2D \cite{cantor2020_prl}), and Cardenas-Barrantes {\it et al.}, in 2D \cite{cardenas2020_pre,cardenas2022_pre} and 3D \cite{cardenas2022_sm}, have stated compaction equations fully determined by the evolution of the microstructure, resolving thus a long lasting issue verified through the very large number of {\it ad hoc} models that have been proposed over the last few decades \cite{Walker1923_The,Carroll1984,Secondi2002_Modelling,Montes2010,Parilak2017,Nezamabadi2019}.

%Problematique & Questions
Still, there is no experimental validation, especially in 3D, of all these advances. Even more important, no clear strategies exist to explore highly deformable granular systems far beyond the jammed state, from the particle scale.

More generally, whether it is in 2D or 3D, the measurement of contacts (position and orientation) and forces or stresses in the grains, as well as the description of their evolution, are mostly based on inverse problem methods. 
Close to the jammed state it is always assumed that the contacts obey a Hertz type law \cite{landau1986_book}.
It is then possible to detect the contacts and to extrapolate the resulting interaction forces using different local field measurements such the deformation of the grain \cite{brujic2003_fd,brodu2015_natcom}, a photoelastic \cite{daniels2017_rsi,abed2019_gm} or thermoelastic \cite{jongchansitto2018_em} signal, or raw measurements by digital image correlation (DIC) \cite{marteau2017_gm}. These inverse methods have also been successfully extended to 2D non-circular grains assemblies \cite{Wang_force_exp2021,Fazelpour_exp_polyg2022} and also to 3D analysis using X-ray tomography with different granular materials \cite{Saadatfar_3D_force_exp2012}.
Careful image analysis work also allows some inroads into the problem \cite{wiebicke2017metrology}.

Far from the jammed point, at least two challenging issues appear: ($i$) Hertz's law is no longer valid, and ($ii$) it is difficult to follow the shape of the particles while detecting new contacts. This last point is particularly exacerbated in 3D and at large packing fraction since it requires high image resolution. Moreover, it is not always possible to use a direct optical approach to measure local particle properties, and non trivial tomographic reconstructions may be necessary.
Most of the 3D experimental studies based on highly deformable, squishy, grains are macroscopic and focus on general properties such as stress release \cite{Indrarantna2019_Use,Tsoi2011_Mechanical}, seismic isolation \cite{Tsang2008_Seismic,Tsiavos2019_A}, or foundation damping \cite{Mashiri2015_Shear}. 
%EA : note: il faut vraiment vérifier cette affirmation, je crois oui?
To the best of our knowledge, only the work of S. Mukhopadhyay and J. Peixinho in 2011 \cite{mukhopadhyay2011_pre}, using fluorescent hydrogel spheres, associated with a tomographic reconstruction, was able to follow the local evolution of a compressed packing up to a density of $0.85$. In particular, they described the evolution of particle connectivity, but unlike what is obtained numerically, they do not report power-law dependencies between the coordination number and packing fraction.

%Reponse
In this work, we present a new 3D experimental technique that measures the  displacement field directly at the grain scale (within grains), without any material or geometrical assumptions. This is performed by combining advanced Digital Volume Correlation (DVC) methods together with accurate x-ray tomography imaging. This method is applied to the paradigmatic case of axial compression with prevented radial strain: a cylindrical sample containing $60$ bidisperse silicone beads of diameter $3$ and $4$~mm is compressed vertically. Micrometer scale glass beads are trapped in the particle material to form a 3D homogeneous random pattern to facilitate DVC. The granular medium is quasistatically compressed in an x-ray scanner to obtain density images of the evolving system, halting loading during scans. Experiments have been repeated four times.
The major outcome of this work is the simultaneous measurement of the topological (contacts and shapes) and bulk properties (internal strain fields) of each grain throughout the whole compression process up to very high packing fraction. This permits the inferrence of the grain-scale stress tensors from the loading history and the material rheology. 
In a context where 3D simulations are still rare and highly demanding in terms of time, our method allows the description of the microstructural and micromechanical properties of any squishy granular system deep in the jammed state. This allows the validation and extension to uniaxial compression constitutive laws recently introduced numerically and in 2D configurations \cite{cantor2020_prl,cardenas2022_sm}, in view of theoretical modeling.

\begin{figure}
\centering
\includegraphics[width=0.7\linewidth]{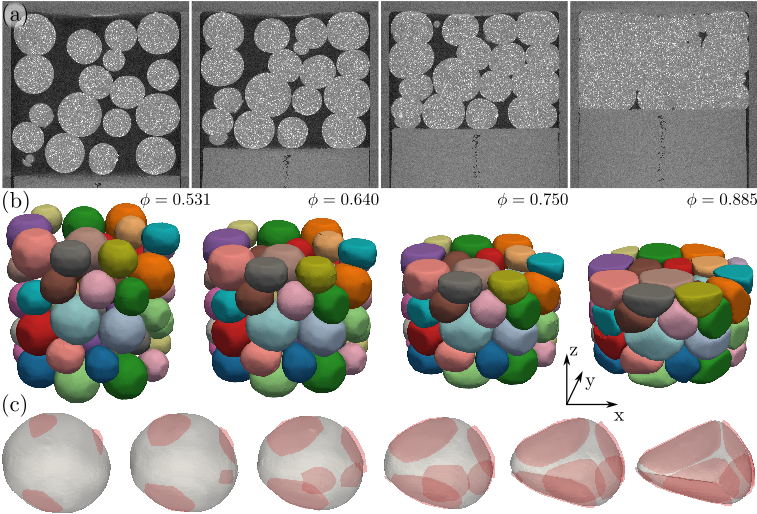}
\caption{Raw and post-processed data: (a): Middle slice of the x-ray scanner density matrix for different levels of compression. The system is compressed vertically from the bottom. (b): Corresponding 3D reconstruction of the particle surfaces from the DVC results obtained from the x-ray scans. (c): Evolution of the shape and contact surfaces (in red) of a given particle when compressed. 
}
\label{fig_0}
\end{figure}

%/<|>\%/<|>\%/<|>\%/<|>\%/<|>\%/<|>\%/<|>\%/<|>\%/<|>\%/<|>\%/<|>\%/<|>\%
% results
\section{Results}

%/<|>\%/<|>\%/<|>\%
\subsection{Compaction and connectivity evolution laws}
\label{Sec_epsilon_Z_Phi_P}

Figure \ref{fig_0}a and b shows tomographic slices and corresponding 3D reconstructions of one of our samples for different levels of compression: from an initially jammed state to a highly compact one. During the compression process the voids within the assembly are progressively filled due to the deformation of the grains. Meanwhile, the packing fraction $\phi$ increases from $\phi_0 \approx 0.53$ (jammed state) to $\phi_{max} \approx 0.9$ for the most compact state obtained. 

\begin{figure}
\centering
\includegraphics[width=0.7\linewidth]{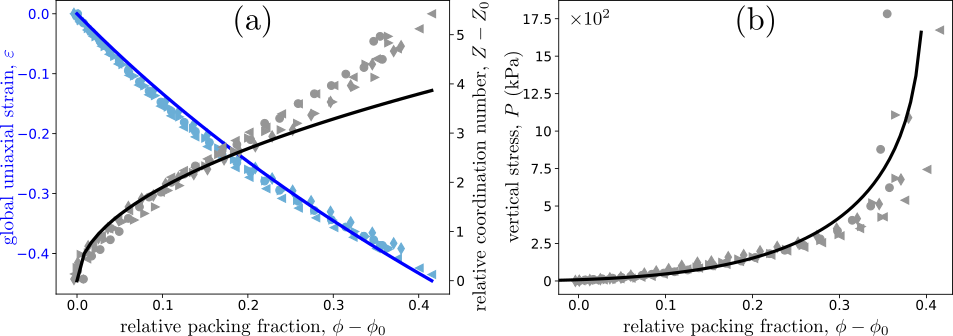}
\caption{Macroscopic evolution. (a): Evolution of the global uniaxial strain, $\varepsilon$, (in blue) and of the coordination number increase from the jamming point, $Z-Z_0$, (in black) as a function of the relative packing fraction, $\phi-\phi_0$. Points are experimental data (different symbols stand for different experiments) while plain lines are analytical models. The line for global strain is a scaling of $-\ln\left( {\phi}/{\phi_0} \right)$, while the black line is a plot of Eq.\ref{eq_zphi}. (b) Compaction curve: applied vertical stress, $P$ as a function of $\phi-\phi_0$. Points are experimental data while the plain line is the analytical model given by Eq.\ref{eq_comp}.}
\label{fig_1}
\end{figure}

In Fig. \ref{fig_1}, deep in the jammed state, we present the evolution of the global vertical strain $\varepsilon = -\ln (H/H_0)$, with $H_0$ the initial height of the cylinder (a, blue curves), the coordination number $Z$ (a, black curves), and the applied vertical stress $P$ (b), as a function of the packing fraction $\phi$.  Basically, $\varepsilon$ decreases as $\phi$ increases following, with a good approximation, a logarithmic law: $\varepsilon \sim -\ln (\phi/\phi_0)$.
The evolution of $Z$ is such that it increases in a non-linear way with $\phi$ from $Z_0 \approx 4.5$ to $Z_{max} \approx 8.5$, as $\phi$ goes from $\phi_0$ to $\phi_{max}$. As shown in the previous 2D numerical/experimental works \cite{durian1995_prl,majmudar2007_prl,katgert2010_epl,vu2019_pre,cantor2020_prl} and recently in 3D \cite{cardenas2022_sm}, this increase follows a power law:
\begin{equation} \label{eq_zphi}
    Z - Z_0 = \xi \sqrt{\phi - \phi_0},
\end{equation}
where $\xi \approx 6$ is fitted from experimental data. The same power-law is observed in our 3D experiments (Fig.\ref{fig_1}(a), black solid line). The small variations from the prediction are due to the experimental uncertainty in the contact detection limited by the image resolution. At large packing fraction we note that the system most likely enters a different regime. Thus, and to the best of our knowledge, our experimental results confirm for the first time the validity of the phenomenological law between $Z$ and $\phi$ up to a certain density. This law has been often verified numerically or in 2D model experiments only \cite{durian1995_prl,majmudar2007_prl,katgert2010_epl,vu2019_pre,cantor2020_prl,cardenas2022_sm}. 

Along with the evolution of $Z$, the compaction of the assembly is also characterised by the $P$ \textit{vs.} $\phi$ curves, shown in Fig. \ref{fig_1}b. We find that $P$ slowly increases with $\phi$ at low density, and then diverges as the packing fraction tends to $\phi^*_{max} \approx 1$, the ``theoretical'' maximum packing fraction. This divergence is expected since the assembly of soft grains begins to behave as a rigid body when the grains almost fill all the voids.
By combining the micromechanical formulation of the granular stress tensor \cite{Andreotti2013}, its limit to small deformations and the evolution of the particle connectivity, we deduce the following equation for the compaction curve $P$ \textit{vs.} $\phi-\phi_0$ (see details in the Methods section):
\begin{equation} \label{eq_comp}
P = E^* \frac{2 (1+2\mu_M)}{3 \pi \Gamma^{3/2}} \frac{\phi^*_{max}-\phi_0}{\phi_0^{3/2}}  Z \phi (\phi-\phi_0)^{1/2} \ln\left( \frac{\phi^*_{max}-\phi_0}{\phi^*_{max}-\phi} \right)
\end{equation}
Where $E^* = E/2(1-\nu^2)$ is the reduced Young modulus, $\mu_M=0.25$ is related to the macroscopic friction of an assembly of spherical grains, $\Gamma=3.89$ is a structural parameter linearly relating the macroscopic deformation and the mean contact deflection, and $Z$ is the coordination number given by Eq.\ref{eq_zphi}.
The power of the proposed compaction law is that it is based only on physical parameters that are easy to measure experimentally. As we can see, our data are very well fitted by Eq.\ref{eq_comp}. This validates and extends to the case of 3D uniaxial compression a recently introduced micromechanical framework allowing to design compaction laws for different loading conditions and material properties \cite{cantor2020_prl}. 

%/<|>\%/<|>\%/<|>\%
\subsection{Quantitative description of grain shapes evolution}

\begin{figure}
\centering
\includegraphics[width=0.9\linewidth]{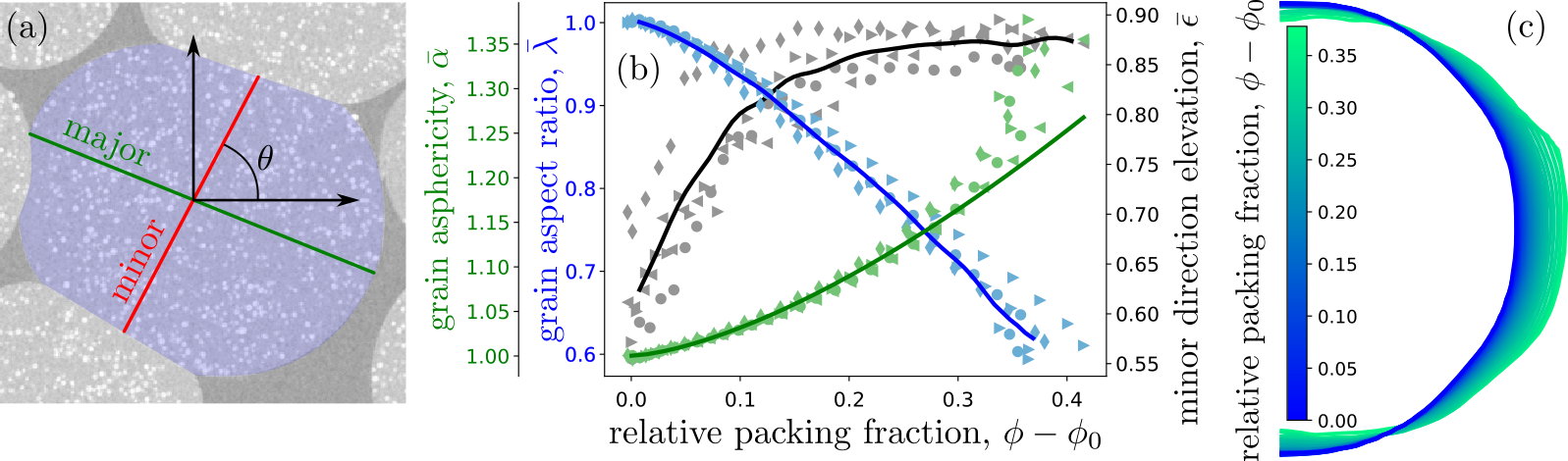}
\caption{Grain shape evolution: (a) Schematic view in 2D of the major and minor directions of a particle and their orientation with respect to the vertical direction. $\theta$ is the elevation angle of the minor direction. (b) Evolution of the mean grain aspect ratio, $\bar{\lambda}$ (in blue), of the average of the sinus of the minor direction elevation, $\bar{\epsilon}_p$ (in black) and of the average grain asphericity, $\bar{\alpha}$ (green) as a function of the packing fraction relative to the jamming point, $\phi-\phi_0$. Points are experimental data (different symbols stand for different experiments). Blue and black plain lines are averaged curves and the green one is the power-law fit: $\bar{\alpha} = 1+ (\phi-\phi_0)^{3/2}$. (c) Evolution of the average shape of the grains as a function of $\phi-\phi_0$. Curves are circumferentially averaged  over each grain of the packing.
}
\label{fig_2}
\end{figure}

The strength of our experimental setup and data post-processing is to obtain precise information at the grain scale, such as the evolution of their shape as well as the evolution of the contact geometries, when the whole assembly is compacted. For example, Fig.\ref{fig_0}c shows the evolution of the shape of one of the grains located in the center of the system, which progressively shifts from spherical to flattened polyhedral shape, with increasing contact areas.

We define the asphericity parameter by $\alpha = \frac{a_p^{3/2}}{3\sqrt{4 \pi}}$, where $a_p$ and $v_p$ are the surface area and volume, respectively, of a given particle $p$. By construction, $\alpha$ equals one for a sphere and is higher for any other geometry. $\bar{\alpha}$ is the average of this quantity over all the grains of the packing. 
Figure \ref{fig_2}(b - green curves) displays the evolution of $\bar{\alpha}$ as a function of the excess of packing fraction $\phi-\phi_0$. We find that the asphericity increases as a power law: $\bar{\alpha} = 1+(\phi-\phi_0)^\beta$, with $\beta=3/2$ and $1$ corresponding to the initial asphericity. Interestingly, a similar trend has recently been observed in 2D/3D numerical simulations of compaction of soft disks/spheres assemblies \cite{cardenas2022_pre,cardenas2022_sm}. Thus, our experiments confirm this seemingly universal geometric feature of soft particles compaction.

Above the loss of sphericity it is important to analyse how particles deform. To do so, for each grain, we consider the evolution of its aspect ratio $\lambda$ and its average orientation $\epsilon$. $\lambda$ is defined as the ratio of the shortest over the longest straight lines joining opposite boundaries of a given grain and crossing its center of mass. $\epsilon$ is the sinus of $\theta$, the elevation angle of the shortest direction (see in Fig.\ref{fig_2}a). As observed in Fig.\ref{fig_2}b, the average particle aspect ratio $\bar{\lambda}$ decreases with $\phi$ from $1$ (spherical shape) to values close to $0.6$. Meanwhile, their average orientation $\bar{\epsilon}$ increases from $0.5$ (anisotropic) to $1$ (horizontally expanded) for $\phi \gtrapprox 0.72$.
From these observations, a schematic picture emerges to describe the compaction from a grain-scale perspective. The grains elongate in the surrounding voids mainly by increasing the largest distance, contributing to the creation of new contacts aligned along the compression direction. In the other directions, the mean grain aspect ratio also increases by shrinking their shortest length. This is also well evidenced in Fig.\ref{fig_2}(c), which shows the average circular profile of the grains. At first this profile is perfectly semi-circular and it then gets progressively larger in width and thinner in height.

%/<|>\%/<|>\%/<|>\%0
\subsection{Linking contact geometry evolution to packing fraction}

\begin{figure}
\centering
\includegraphics[width=1\linewidth]{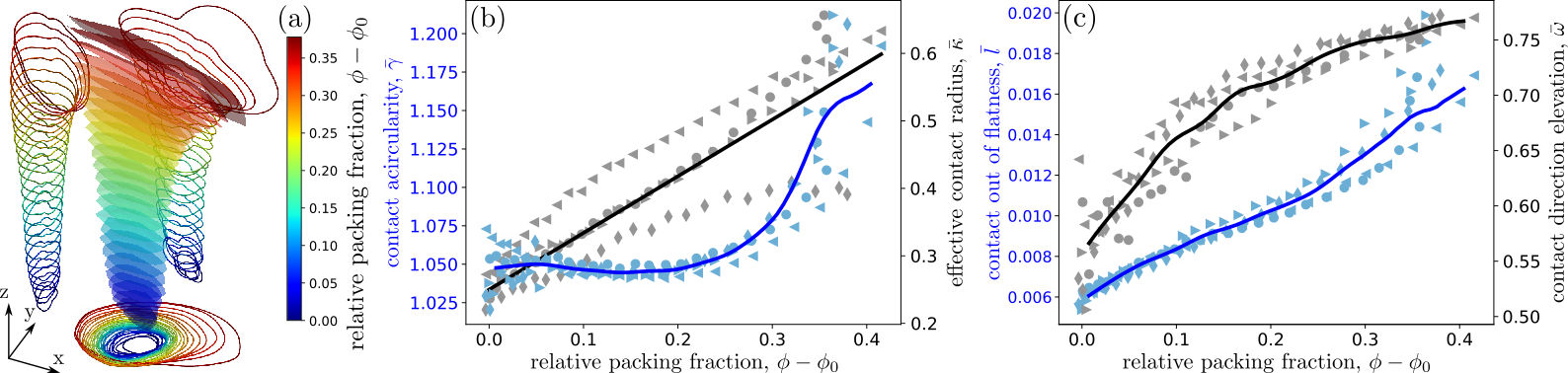}
\caption{Contact geometry evolution. (a): Evolution of the shape of a given contact when the system is compressed. Projections of the contact contours on three orthogonal directions are also shown. (b,c): Evolution of the average contact acircularity, $\bar{\lambda}$ (a-blue), the average effective contact radius, $\bar{\kappa}$ (a-black), the average contact non-planarity, $\bar{l}$ (b-blue) and the average contact direction elevation, $\bar{\omega}$ (b-black) as a function of the distance of the packing fraction from the jamming point, $\phi-\phi_0$. 
Points are experimental data (different symbols stand for different experiments) while plain lines are averaged curves except for $\bar{\kappa}$ where it is a linear fit with slope $2.94 \pm 0.03$.\\ }
\label{fig_3}
\end{figure}

Along with the grain shape, the evolution of the contact surface geometries between grains is also probed during the compression. As an illustration, Fig.\ref{fig_3}a displays the evolution of a typical contact surface within the sample. Qualitatively, close to $\phi_0$ the shape of a contact can be considered as circular. However, as the compression goes deeper into the jammed state, the contact surface gets larger and exhibits a non-trivial concave shape. As for the grain shape, for each contact, we define the mean contact surface acircularity $\gamma =  \frac{p_c^2}{4\pi s_c}$, where $q_c$ and $s_c$ are the perimeter and the surface of this contact, respectively. $\bar{\gamma}$ is the average of this quantity over all the contacts of the packing.
In Fig.\ref{fig_3}b the blue curve shows the evolution of $\bar{\gamma}$, as a function of the excess packing fraction $\phi-\phi_0$.
We find that, on average, the contact surfaces can be considered as circular up to fairly high packing fraction, close to $0.83$.
Before this density $\bar{\gamma} \approx 1$, but beyond, $\bar{\gamma}$ increases suddenly to values close to $1.2$.
This highlights the strong deviation of the contact surfaces from a circular shape. The shape of this curve is reminiscent of the divergence of $P$ observed in Fig.\ref{fig_1}b. Along with $\gamma$, the contact geometry is also characterised by the mean curvature of the contact surfaces $l$.
This latter is defined by the contact non-planarity, the mean distance between the real contact surface and its average plane. This quantity is then normalised by the mean particle radius $R_p$. The evolution of its average value $\bar{l}$ as a function of the excess of packing fraction $\phi-\phi_0$ is shown in Fig.\ref{fig_3}c.
We observe that $\bar{l}$ slightly increases with $\phi$ and remains overall less than $1\%$ of the mean grain radius until, as for $\bar{\gamma}$, a cross-over is observed at $\phi \approx 0.83$. Then, a rapid increase is observed and $\bar{l}$ exceeds more than $1\%$ of the mean grain radius.
In other words, from a certain level of compression, the grains indent each other and the contact surfaces are no longer symmetrical. This leads to concave-shaped contacts as observed in Fig.\ref{fig_3}a.

To quantify the evolution of the size of the contact, we define the relative effective contact radius $\kappa = \frac{\sqrt{s_c/\pi}}{R_p}$. Using a mean field approach that combines the geometrical properties of an Hertzian contact together with the $Z$ \textit{vs.} $\phi$ relation (Eq.\ref{eq_zphi}), we obtain (See Methods Section):
\begin{equation} \label{eq_Rphi}
    \bar{\kappa} \propto \phi - \phi_0.
\end{equation}
We see in Fig.\ref{fig_3}b that the linear dependency predicted by Eq.\ref{eq_Rphi} is well satisfied up to large $\phi$ values deep in the jammed state. However, from $\phi \approx 0.83$, the data slightly deviate from this prediction since the circularity and flatness assumptions of the contact surface are not valid anymore. Going further in the contact analysis, in Fig.\ref{fig_3}b we show the evolution of the mean contact direction $\bar{\omega}=\langle \sin \theta_c\rangle_c$, where $\theta_c$ is the elevation angle of the normal angle to the contacts. We find that $\bar{\omega}$ increases from $0.5$ (anisotropy) at $\phi_0$ to $\sim0.75$ (mainly vertical) for $\phi\rightarrow \phi_{max}
^*$. %$\phi \approx \phi_{max}$. 
This indicates that the contact surfaces are preferentially oriented perpendicular to the direction of compression as the packing fraction is increased. This is consistent with the preferential direction of grain elongation discussed in the previous section.

 %In fig. \ref{fig_3}-b we show the evolution of a typical contact. This latter gets bigger while its overall shape is less and less circular and its envelop is more and more concave. 
 %These qualitative observations are validated by fig.\ref{fig_3}-a. This figure shows the variation of the the average contact acircularity, $\bar{\xi}$ when the system is progressively compressed, with $\xi=frac{p^2}{4 \pi s}/$, $p$ the contact perimeter and $s$, its area. 
 %This quantity first increases almost linearly and then more abruptly when the system is highly compacted. 
 %This figure also shows the evolution of the average out-of-flatness of each contact, $o$. This latter quantity is the volume included between a contact shell and its average plane. 
 %On average, $o$ is constant at the beginning of the loading process and rapidly increases when the system is highly compact. This cross-over for $\phi \approx 0.8$ corresponds with the one observed for $\bar{\xi}$. This means that at a certain level of compression in the contact process one particle indent the other one, the contact is not symmetric anymore. 
 %Also as shown by the evolution of the average contact direction elevations, $c=sin(\theta_c)$, contacts are first isotropic and then tends to align in the compression direction. $\theta_c$ is defined just like $\theta$ shown in fig.\ref{fig_2}-a for contact orientation.

%/<|>\%/<|>\%/<|>\%
\subsection{Deeper within the grains: strain and energy density fields}
\label{Sec_Energy}
\begin{figure}
\centering
\includegraphics[width=0.5\linewidth]{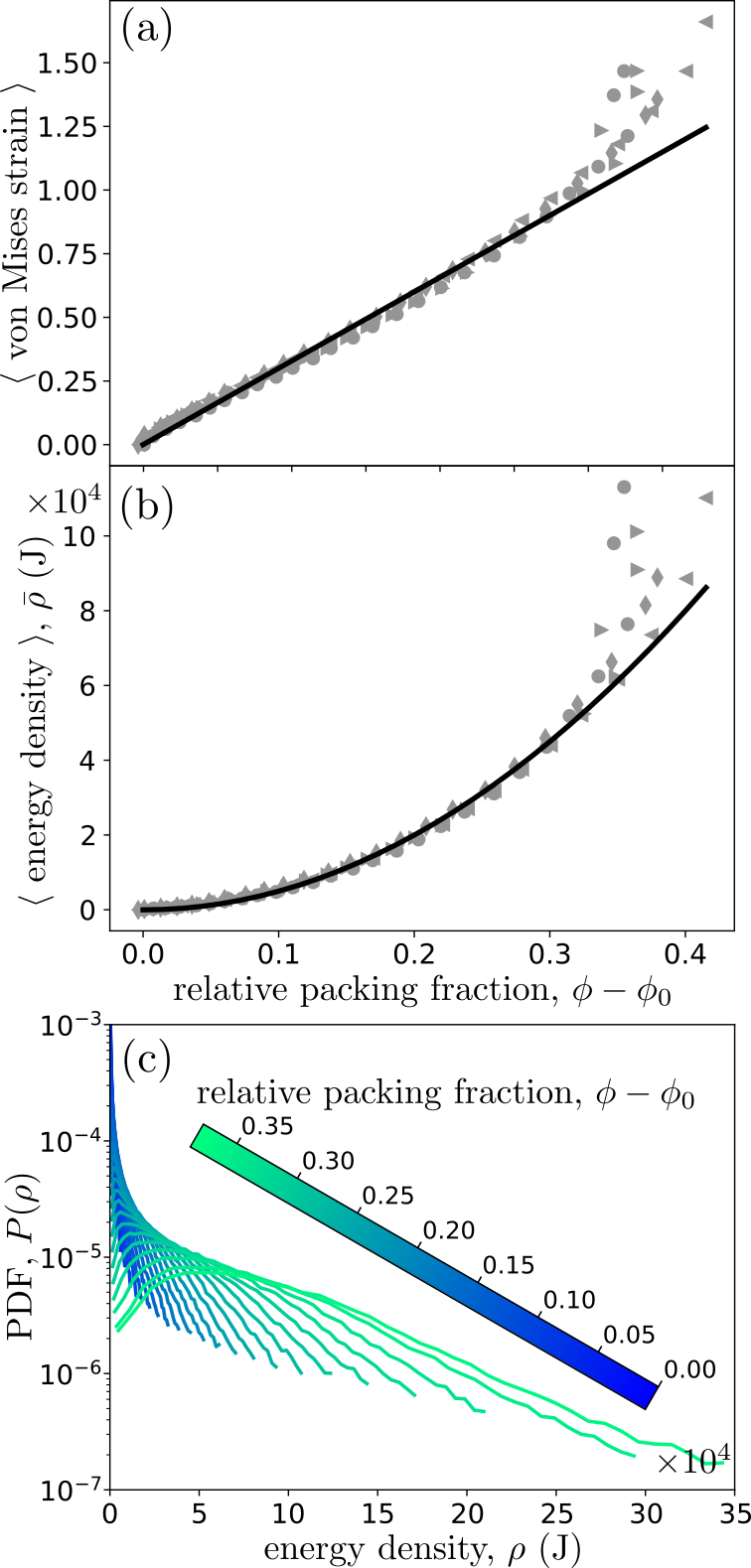}
\caption{Local mechanical evolution. Evolution of the average von Mises strain (a) and of the average energy density, $\bar{\rho}$ (b) as a function of the distance of the packing fraction from the jamming point, $\phi-\phi_0$. Points are experimental data (different symbols stand for different experiments) while in (a) the line show a linear fit with slope $3 \pm 0.1$ and in (b) plain curve is the quadratic fit $5\times10^{5} (\phi-\phi_0)^2$. (c) Evolution of the probability density function (PDF) of the energy density, $P(\rho)$ when a system is compressed.
}
\label{fig_4}
\end{figure}

The presented method not only allows the measurement of the 3D evolution of grain boundaries, but also provides the local displacement field inside each grain, even deep in the jammed state, for highly deformed grains. The right Cauchy-Green deformation \cite{taber2004_book} tensor and its von Mises norms are deduced from this displacement field \cite{vu2019_pre_bis}. Also, assuming an hyperelastic behavior of the silicone used to make the beads \cite{vu2019_pre_bis,vu2019_pre,vu2019_em}, we compute the energy density stored in the material \cite{taber2004_book,vu2019_pre}, $\rho$. Figure \ref{fig_4}(a) shows the evolution of the mean von Mises strain when the granular system is loaded. We find that the mean von Mises strain increases first linearly with the packing fraction. On average, this implies a proportionality relation between the local and the global strain applied to the system. 
%\ea{-- Ici il n'y a pas un lien a faire avec la prediction de Rc avec phi mais aussi de P avec Phi ou quelque part on utilise que epsilon egale delta ? Si oui, il faut le mentionner dans la partie methode.}
Still for higher packing fraction, from $\phi \approx 0.83$ the local strain enters a regime where it increases more rapidly. In fig.\ref{fig_3}(b), we show the evolution of the average energy density when the system is compressed. This latter quantity increases quadratically with the packing fraction. This is a direct consequence of the linear relation between the strain and the packing fraction. %\ea{-- cette remarque je suis pas sur de comprendre, on a montré que strain et packing fration c'était un logartime tout au début? } \jb{oui mais à l'ordre 1 ça devient linéaire non?}
It implies that even for quite high packing fractions, from a coarse-grained point of view, the system behaves as a solid. Still, when looking at the statistical distribution of the energy field, presented in fig.\ref{fig_4}(c), this analogy is no longer true. Indeed, the probability density function (PDF) of $\rho$ is Gaussian-like for gently squeezed system but for more compressed systems, it displays an exponential tail reminiscent of what is observed for the PDF of the interaction forces between rigid grains \cite{radjai1996_prl}. 

%/<|>\%/<|>\%/<|>\%/<|>\%/<|>\%/<|>\%/<|>\%/<|>\%/<|>\%/<|>\%/<|>\%/<|>\%
% discussion
\section{Discussion}

We have overcome one of the main experimental challenges recently identified as such by the granular matter community \cite{bares2022_pp,schroter2022_pp,manning2023_prl}. The novel experimental framework we present in this paper permits to travel from macro to micro observables --and conversely-- of a highly squeezed granular system deep in the jammed state up to very large packing fraction. The strain tensor in the material of each particle of the whole system is recorded during the compression process permitting to compute the evolution of the corresponding stress tensor and energy density. The change in shape of the particles and their contacts are deduced and the classical --strain \textit{vs.} stress-- measurements are also monitored. No classical mechanics observables of the system are kept unknown during the whole loading process opening the way to a thorough analysis of this complex multi-scale mechanical process. 

This novel set of information permitted first to check that the phenomenological exponential law between the system coordination number, $Z$, and packing fraction, $\phi$,  still holds in 3D. 
So far this had been only observed for 2D experiments \cite{durian1995_prl,katgert2010_epl,majmudar2007_prl,vu2019_pre}. 
%Still our data shows that at very high packing fraction, so far unobserved \ea{experimentally}, the evolution regime change and turns linear, the number of grain-grain contact increases linearly with the system density. --> je propose d'enlever ca. On dit dans la section II.A que la deviation a grande compacité c'est peut etre due aux erreur de mesures. Dans l'article 3D numérique de Manuel ca marche bien jusqu'au bout. De plus, en laissant la loi en puissance dans le model de compaction ca marche bien.
This point extends the domain of validity of this relation and suggests that more contacts are created than what was initially expected. 
The compaction law linking the pressure applied to the system and its packing fraction, built for 2D systems and in 3D for isotropic \ea{compression} %compaction 
have been extended to the case of uniaxially loaded 3D systems. The strength of this theoretical framework, derived from the micro-mechanical expression of the granular stress tensor, is to be based only on physical parameters that are easy to measure experimentally. Our experimental results outstandingly validate this law even deep in the jammed state. This would permit to predict with great precision the behaviour of a loaded system just by knowing the exact nature of the grains.

%Extending what have been observed in 2D \cite{cardenas2022_pre}, EA--> Je propose d'enlever ca pour pas donner des arguments que c'est "juste" une extension, 'est vraiment plus que ca.
Moreover, the results presented in these series of 3D experiments shows that the mean contact geometry varies linearly with the packing fraction. Just like for the compaction law, this relation is supported by a mean field approximation developed in the small deformation framework. It is remarkable that even deep in the jammed state, where grains are extremely deformed through multiple contacts and that the Hertz contact hypothesis are highly violated \cite{landau1986_book,vu2019_em}, the mean-field analytical framework still holds and gives greatly reliable results. This point is good omen to find a unified rheological law describing the mechanical behaviour of a system from the jammed state up to bulk material whatever the loading process.

Still by tracking the local strain energy during compression, we reveal that, for very high packing fraction, a transition from granular-like to continuous-like material exists. For a packing fraction close to $0.83$, the average energy density exits a quadratic law with the packing fraction and the average strain exits a linear law. The description of these quantities remains complex to catch analytically in these high density regimes. They correspond to regimes where particles are highly distorted to fill the last remaining porosities. This is is agreement with the fact that above $\phi \approx 0.83$, the average particle asphericity also exits a well defined power-law behaviour with the packing fraction. Below this threshold the material remains remarkably well described by the material rheology taking granular structure into account; on the contrary, above, it enters a more extreme regime where observables vary abruptly. This point, that has been also observed in 2D \cite{vu2019_pre_bis,cardenas2022_pre}, clearly evidences a sharp crossover in the material behaviour.  

Our experimental framework evidences heuristic and analytical laws linking different observables of densely packed granular systems. This provides insights into the granular micro-mechanisms and how they are responsible for a specific macroscopic behavior. On top of permitting to predict the evolution of a densely packed system, our findings permit to deduce the global state of a system through local observations --and conversely. This opens new perspectives for rheological analysis of highly deformable grain assemblies in various fields ranging from biology \cite{czajkowski2018_sm,yan2019_prx,manning2023_prl}, engineering \cite{cooper1962_jacs} and geophysics \cite{tsang2008_eesd,tsiavos2019_sdee}, to name just a few.

%/<|>\%/<|>\%/<|>\%/<|>\%/<|>\%/<|>\%/<|>\%/<|>\%/<|>\%/<|>\%/<|>\%/<|>\%
% method
\section{Methods} 

\begin{figure}
\centering
\includegraphics[width=1\linewidth]{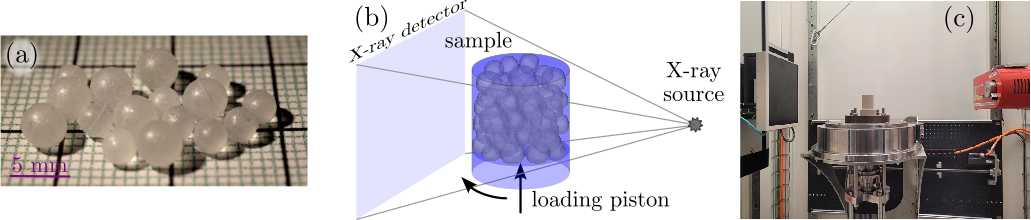}
\caption{(a): picture of the custom-made silicone particles. The large particles are $4$~mm wide, while the small ones have a diameter of $3$~mm. (b): schematic view of the in-situ uniaxial compression set-up. (c) picture of the experimental set-up.}
\label{fig_M1}
\end{figure}

%/<|>\%/<|>\%/<|>\%
\subsection{Sample preparation and {\it in-situ} compression}

The granular samples are made of $60$ silicone beads, $24$ of diameter $4$~mm and $36$ of diameter $3$~mm (see fig.\ref{fig_M1}-a). These particles are molded with SortaClear 18 silicone by SmoothOn company \cite{silicone}, mixed with $35$\% in mass of glass beads of diameter between $40$ and $70$ $\mu$m. The silicone is viscous enough for the glass beads not to sediment during polymerisation. Just before molding, the silicone is degased to remove air bubbles. Each grain is coated with talc powder to avoid the particles to stick together. They end up having a friction coefficient of $0.59 \pm 0.15$. The Young modulus of the material is $E = 0.663$~MPa and its Poisson ratio is $\nu = 0.289$.
\\

To account for the variability of the initial state, $4$ independent experiments are performed.
Each specimen is prepared by placing particles individually and layer by layer inside an oedometer of \SI{15}{\milli \meter} in diameter.
To avoid segregation and crystallisation effects, each ``layer'' consists of $2$ large and $3$ small particles. The particles are deposited until the height of the assembly reaches \SI{15}{\milli \meter}, yielding a specimen aspect ratio of 1.
The oedometer is then closed and mounted on top of the rotation stage of the x-ray tomography cabin, built by RX-Solutions \cite{viggiani2015_gtj,karatza2019_gm}.
The loading system is located below the setup, thus the loading is applied from the lower plate of the oedometer, as shown in Fig.\ref{fig_M1}-c.
The oedometric compression is performed ``in-situ'', \textit{i.e.}, imaging the specimen while performing the test step-by-step: the specimen is compressed at a loading speed of \SI{4}{\micro \meter}, using displacement steps of \SI{250}{\micro \meter}, after which, the loading system is paused and the specimen is imaged.
To perform a scan, $3$ x-ray projections are averaged for each of the 1120 different radial positions (see Fig.\ref{fig_M1}-a), and a 3D field is reconstructed using the filtered back-projection algorithm proposed by \cite{feldkamp1984practical}, as implemented within the XAct software provided by RX-Solutions, saved on a 16-bit data range.
The x-ray tomography scans are performed using a pixel size of \SI{15}{\micro \meter \per pixel}, resulting in more than $200$ pixels per particle diameter.
The loading force is measured externally, and each specimen is compressed until a packing fraction of $\phi \approx 1$ is reached.

\begin{figure}
\centering
\includegraphics[width=1\linewidth]{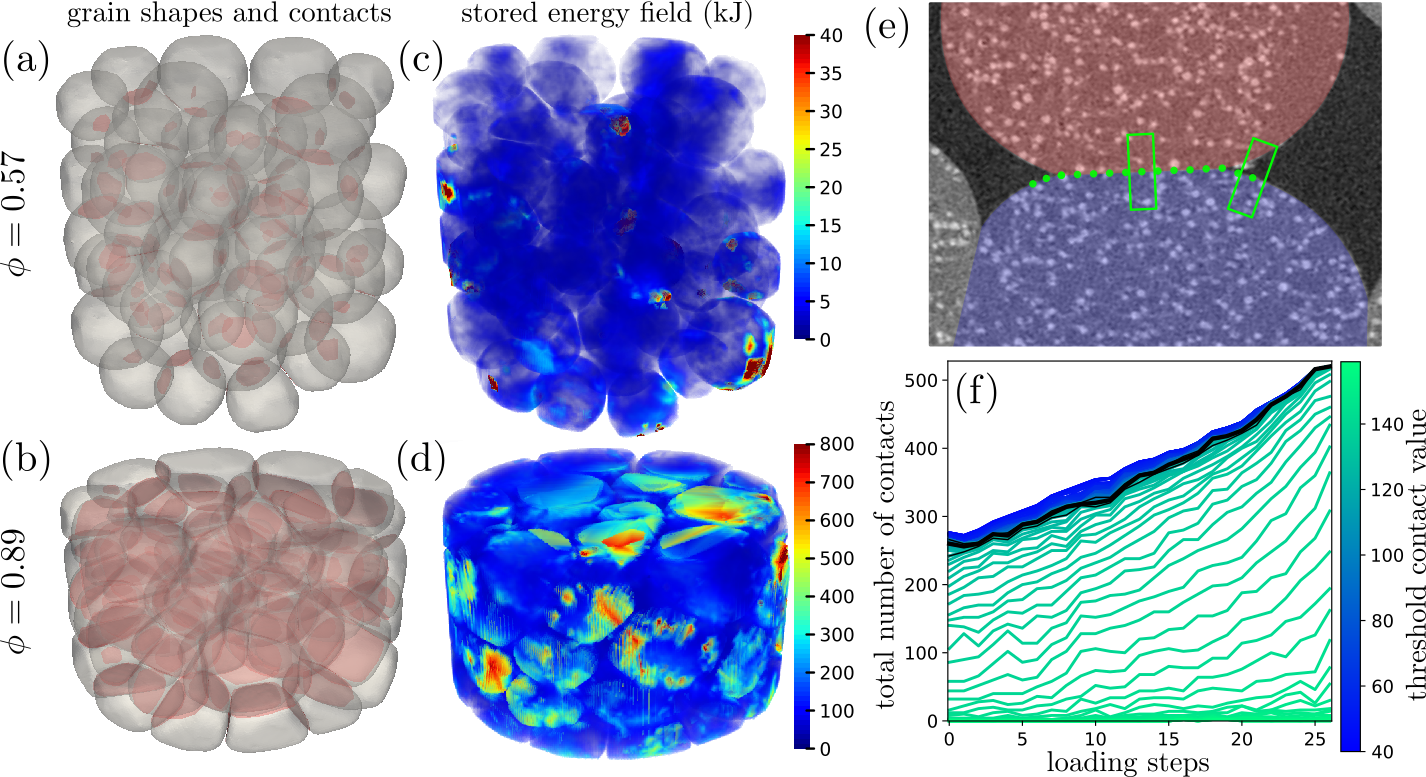}
\caption{(a)-(b): evolution of 3d contours of the particles as obtained from DVC. The contact areas are shown in red. (c)-(d) corresponding packing showing stored energy density fields. (e): 2D view of a x-ray slice zoomed on the contact between two grains. Green dots are potential contact points where the average x-ray density is probed inside the green rectangles. The average value of this density in these subzones constitutes a contact detection threshold. (f) evolution of the number of contacts in the whole system by varying the threshold on the contact criterion.}
\label{fig_M2}
\end{figure}

\subsection{Numerical method for post-processing}
From each experiments we obtain $\sim 25$ $16$~bit density matrices of dimension $1200 \times 1200 \times 1800$. We first convert the matrices into $8$~bit. Then we isolate the large cluster of connected grains from the rest of the system on the initial undeformed image by means of thresholding. Eroding sequentially this cluster, we isolate the centers of the grains. To segment grains in the initial image we detect voxels on the borders of this large cluster containing all the grains. Considering these voxels one by one we check if the lines going from it to the two closest particle centers stay included in the cluster. If so, that means the voxel belongs to a contact border. These voxels forming penny shaped clusters give the border of the contact, where we have to separate particles. We look for the plane in which each of these clusters are included and cut the big cluster of grains along these planes. As shown in the bottom left part of fig.\ref{fig_0} we obtain the clusters of voxels covering each particle.

From the initial particle clustering, regular cubic meshes are built inside each grain. They form a network of correlation cell centers that will be tracked from one image to the next all along the compression. The distance between the nodes of this network is $12$~vx. From one image to the next, the solid rigid motion of each grain is first computed: on the first image, around the closest correlation cell center from the particle center of mass, a cubic sub-matrix of side $36$~vx is extracted. By means of Fourier transform it is convolved with the submatrix at the same location in the next image. The position of the maximum of convolution gives the particle translation with a vx accuracy. To improve this displacement measurement, we minimize the squared difference between the first submatrix and the second one deformed by a first order shape function. Once the solid rigid motion of each particle is obtained, by applying the same process as the one described for the particle center of mass to each correlation cell center we obtain the displacement field inside each particle. In this latter computation step, the correlation cell size is $18$~vx and if the correlation criterion is low enough, then the shape function is just a translation. The process is sequentially repeated for each consecutive pair of images. 

From the displacement field obtained from DVC, the displacement of the particle borders is deduced and the evolution of the particle shapes is tracked along the compression process. To measure contact surfaces, we first discretize the border into a network of points separated by $18$~vx. for each of these points, if they are close to another point belonging to another particle we extract a parallelepipedic submatrix centered around this point and aligned perpendicularly from the boundary as shown in fig.\ref{fig_M2}-e. We compute the average value of this submatrix. The higher this density value, the higher the quantity of matter in the submatrix, so the higher the probability of contact. In fig.\ref{fig_M2}-f we have plotted the evolution of the number of contacts for different thresholds on this density value to consider the contact as real. We observe that the density of curves increases for a threshold of $120$, which is close to the mean density ($2^8/2$). This is the threshold we picked.

\subsection{Derivation of the compaction equation for 3D uniaxial geometry}
%EA : 
%ici je reprends dans les grandes lignes ce que l'on a mis dans le SoftMatter, 
%J'ai essayé de condenser au maximum en voulant donner les différentes étapes
%On peut certainement réduire encore.
We briefly recall the main ingredients of the theoretical framework presented in \cite{cardenas2022_sm} 
and we extend it to fit uniaxial 3D geometry.
In this case, $P$ is related to the granular stress tensor through its 
$zz$ component by $P=\sigma_{zz}$. The granular stress tensor is given by $\bm{\sigma}  = n_c \langle \bm{f^c} \otimes \bm{\ell^c} \rangle_c$ \cite{Andreotti2013}, 
where $\bm{\ell^c}$ is the vector between the centers of the grains and $\bm{f^c$} the force vector at a contact $c$. $n_c=N_c/V$ is the density of contact, with $N_c$ the total number of contacts in the volume $V$. Considering a small particle size distribution around the mean diameter $d$ and $V_p = (\pi/6)d^3$ the volume of a sphere $p$, we obtain $n_c \simeq 3Z\phi/(\pi d^3)$, with $Z=2N_c/N_p$ the coordination number.
From the definition of $\sigma_{zz}$ via the principal values of $\bm{\sigma}$, we get:
\begin{equation}\label{eq:Pzz_sigma}
P \simeq \alpha \frac{ \phi Z} {\pi} \sigma_{\ell},
\end{equation}
with $\sigma_{\ell} = \langle f^c \cdot \ell^c \rangle_c/d^3$, a measure of the mean contact stress and $\alpha$ a constant related to various microscopic parameters pertaining to the contact and force network specific of uniaxial compression, but hardly accessible by means of experiments. However, theoretical developments \cite{Ouadfel2001} have shown that these microscopic parameters can be added together to build the shear strength parameter $\mu_{M}$ of the material allowing to write $\alpha\sim1+2\mu_{M}$. For an assembly of frictional spheres $\mu_M$ is close to $0.25$. Note that, Eq. \ref{eq:Pzz_sigma} is mostly known in its simpler form \cite{Agnolin2007c,Andreotti2013}, 
where $\mu_M$ is fixed to $0$ when considering the mean stress from the trace of $\bm \sigma$.

Close to the jammed state ({\it i.e.}, in the case of small deformations), assuming that the contact forces follow a Hertz law and verifying that (as discussed in Sec.\ref{Sec_Energy}) the macroscopic deformation $\varepsilon$ is linearly related to the mean contacts deformation $\langle \delta^c/R_p\rangle_c$, with $\delta^c$ the reduction of radius at a contact $c$ (via a parameter $\Gamma\sim3.8$ measured in our experiments), we can rewrite Eq.\ref{eq:Pzz_sigma} as $P_{SD} = -\alpha\frac{2E^*}{3\pi\Gamma^{3/2}}Z_0\phi\ln^{(3/2)}\left(\frac{\phi}{\phi_0}\right)$.
Now, deep in the jammed state, we rely on the macroscopic hypothesis formulated by Carroll and Kim \cite{Carroll1984}. Using an analogy between the compaction process and the collapse of a cavity within an elastic medium they shown that, necessarily,  $P \propto \ln[(\phi^*_{max}-\phi)/(\phi^*_{max}-\phi_0)]$.
By combining this large deformation approach, together with the small one and Eq.\ref{eq:Pzz_sigma} the compaction equation becomes the one presented in Sec. \ref{Sec_epsilon_Z_Phi_P}.

\subsection{Predicting the linearity between $\bar{l}$ and $\phi$}
First, let's remark that $\bar{l}$ can be rewritten as a mean over all grains and contacts: $\bar{l} \sim Z.r_c/R_p$. Then, since we have verified (see Sec.\ref{Sec_epsilon_Z_Phi_P}) that $(\phi-\phi_0)/\phi_0 \sim \varepsilon$ and also that $\varepsilon \sim \langle \delta^c/R_p\rangle_c$ (see Sec.\ref{Sec_Energy}), we get the following proportionality between $Z$ and $\delta$: $Z \sim \sqrt{\delta/R_p}$. Finally, from the Hertz theory we get that the radius of the contact surface between two spheres touching at a contact $c$ is given by $r_c \sim \sqrt{R^*\delta^c}$, with $R^* \equiv R_p$ the effective radius and $\delta^c$ the reduction of radius at the contact. Injecting the two above proportionalities within the definition of $\bar{l}$ leads to $\bar{l} \sim \sqrt \delta \sqrt\delta \sim \phi-\phi_0$.

\subsection{Data availability} 
The data that support the findings of this study are available from the corresponding author.

\subsection{Code availability} 
The codes that support the findings of this study are available from the corresponding author.

\subsection{Author contributions}
E.An., J.B., E.Az. and M.R. conceived the experiment; J.B. made the samples; G.P. and M.C.-B. performed the experiments; J.B. and G.P. post-processed the data; J.B., M.C.-B., G.P., E.An., E.Az. and M.R. analyzed the data; M.C.-B., E. Az. and M.R. set-up the analytical model; E.Az., J.B., G.P., M.C.-B. drafted the manuscript; all authors reviewed the manuscript.

\subsection{Acknowledgements}
We are grateful to G. Camp and S. Devic for helping with the sample preparation.

\bibliographystyle{plain}
\bibliography{biblio}

\end{document}